\documentstyle[prd,eqsecnum,preprint,aps]{revtex}

\begin{document}
\draft
\title{Recollapse of the Closed Tolman Spacetimes}
\author{Gregory A. Burnett}
\address{Department of Mathematics, North Carolina State University,
Raleigh, NC\ \ 27695-8205}
\date{19 July 1993}
\maketitle
\begin{abstract}
The closed-universe recollapse conjecture is studied for the
spherically symmetric spacetimes. It is proven that there exists an
upper bound to the lengths of timelike curves in any Tolman spacetime
that possesses $S^3$ Cauchy surfaces and whose energy density is
positive.  Furthermore, an explicit bound is constructed from the
initial data for such a spacetime.
\end{abstract}
\pacs{04.20.Cv, 04.20.Jb, 98.80.Hw}
\narrowtext

\section{Introduction} \label{sec:intro}

Must the Universe end?  The {\it closed-universe recollapse
conjecture} offers the intriguing possibility that the ultimate fate
of the Universe may be a consequence of its spatial topology.  The
conjecture says, roughly, that a closed Universe with ordinary matter
must expand from an initial singular state to a maximal size and then
recollapse to a final singular state, thus giving the Universe a
finite lifetime \cite{BGT,BTa,BTb,MT}.  In order to investigate this
conjecture, we formulate a precise version as follows.

First, a restriction on the matter content is necessary for otherwise
counterexamples to any reasonable conjecture are easily constructed.
Exactly what energy condition should be imposed is an open issue.  but
it has proved useful in prior investigations \cite{gaB} to demand that
the dominant-energy and non-negative-pressures conditions hold.  The
dominant-energy condition is the demand that $G_{ab}t^au^b \ge 0$ for
all future-directed timelike $t^a$ and $u^b$.  The
non-negative-pressures condition is the demand that $G_{ab}x^ax^b \ge
0$ for all spacelike $x^a$.  Through Einstein's equation and for a
stress-energy tensor that possesses a timelike eigenvector with
eigenvalue $-\rho$ and principal pressures $p_i$, together these
conditions are equivalent to the inequalities $\rho \ge p_i \ge 0$.
It is interesting to note that simply demanding that the more standard
energy conditions of dominant-energy and timelike-convergence
($R_{ab}t^at^b \ge 0$ for all timelike $t^a$) hold is not sufficient
to guarantee the recollapse of even the $k=+1$
Friedmann-Robertson-Walker spacetimes.  A simple counterexample is
given by the Friedmann-Robertson-Walker spacetime with scale factor
$a(t)=t$ \cite{BGT}.

Second, by a spacetime having a finite lifetime we shall mean that
there is a finite upper bound to the lengths of timelike curves in
that spacetime.  (A closed universe with such a bound is known as a
Wheeler Universe \cite{MT}.)\ \ If the spacetime satisfies the
timelike-convergence condition and a genericity requirement, then the
existence of a maximal hypersurface is sufficient to guarantee that
such an upper bound exists \cite{BTa,MT}.  However, proving the
existence of a maximal hypersurface seems to be a very difficult task
as spacetimes satisfying our energy-condition requirement can be
constructed that don't possess such hypersurfaces.  For example,
consider the spacetime obtained by taking the past of any expanding
spatially homogeneous hypersurface of a $k=+1$
Friedmann-Robertson-Walker spacetime with positive energy density and
pressure.  Although in this case it is possible to continue the
spacetime to the future, examples can be constructed where the
pressure diverges in a finite time (thus preventing a future
extension) with the spacetime always expanding \cite{BGT}.  Note that
in these examples, although the development of a maximal Cauchy
surface is prevented by the ``halting'' of the evolution of the
spacetime, there is a finite upper bound to the lengths of timelike
curves in the spacetime.  In light of this, rather than attempting to
impose conditions designed to guarantee that a maximal hypersurface
can develop, we simply investigate the question of whether there
exists a finite upper bound to the lengths of timelike curves in the
spacetimes under consideration.

Lastly, although the spacetimes being studied may not contain a
maximal hypersurface, the existence of such a surface should not be
precluded because of the spacetime's topology.  For such a maximal
hypersurface $\Sigma$ to exist, it is necessary that the scalar
curvature associated with the induced metric on $\Sigma$ be
non-negative.  (This is easily seen using the initial-value constraint
equation and the non-negative-energy condition.)\ \ However, there are
few compact orientable three-manifolds that admit metrics with
non-negative scalar curvature \cite{dmW}.  We further narrow the
allowed topologies for $\Sigma$ by demanding that the induced metric
be non-flat, thereby excluding such possible counterexamples as the
identified Minkowski spacetime with spatial topology $S^1 \times S^1
\times S^1$.  The resulting allowed Cauchy surface topologies are
$S^3$, $S^1 \times S^2$, and those constructed from these by making
certain identifications and connected summations \cite{BTa,BTb}.

Combining these requirements, we arrive at the following

{\it Conjecture.}
There exists an upper bound to the lengths of timelike curves in any
spacetime that possesses $S^3$ or $S^1 \times S^2$ Cauchy surfaces and
that satisfies the dominant-energy and non-negative-pressures
conditions.

Currently, there is no known counterexample nor proof of this
conjecture.  However, there are results offering evidence for its
truth.  The simplest is the fact that the $S^3$ case of the conjecture
holds for all spatially homogeneous and isotropic spacetimes
\cite{TW,BGT}.  Lin and Wald \cite{LW} have generalized this last
result by relaxing the isotropy assumption.  That is, the above
conjecture holds for the Bianchi type IX spacetimes.

Further, it is known that the $S^1 \times S^2$ case of the above
conjecture holds for spacetimes that are spherically symmetric
\cite{gaB}. The basic idea behind the proof of this result is simple.
Associated with the spherically symmetric spacetimes are two scalar
fields $r$ and $m$.  The field $r$ is simply the usual ``size''
associated with the spheres of symmetry while $m$ is a ``quasi-local
mass'' associated with these spheres. (These are discussed in detail
in Sec.~\ref{sec:basics} below.)\ \ In the $S^1 \times S^2$ case it
turns out that $r$ is positive and globally bounded from above and $m$
is globally bounded from below by a positive constant.  Further, as
$r$ changes as a function of proper time $t$ along any timelike
geodesic, it must obey the inequality $d^2r/dt^2 \le - m / r^2$.
These facts together allow us to easily conclude that there is a
global finite upper bound to the lengths of timelike curves in these
spacetimes.  (Further details of this argument can be found in
Ref.~\cite{gaB}.)

However, it is unknown whether the $S^3$ case of the above conjecture
holds for spacetimes that are spherically symmetric.  In the $S^3$
case we still find that $r$ is globally bounded from above but now we
know only that $m$ is non-negative.  Because of this we are unable to
obtain in such a simple manner an upper bound to the lengths of
timelike curves.

Herein, we show that the problems in proving the above conjecture for
the spherically symmetric spacetimes with $S^3$ Cauchy surfaces can be
overcome for the nowhere empty Tolman spacetimes.  These are the
spherically symmetric spacetimes whose matter content is dust (a
perfect fluid with vanishing pressure) \cite{LL}.  The method of proof
can be understood roughly as follows.  First, we introduce a
``time-function'' $t$ whose boundedness entails an upper bound to the
lengths of timelike curves.  Next, we divide the spacetime into three
regions: a ``northern'' region, a ``middle'' region, and a
``southern'' region.  Using the upper bound for $r$ and the positivity
of $m$ in the ``middle'' region and on the boundaries of the
``northern'' and ``southern'' regions, the boundedness of $t$ in these
regions follows in a straightforward manner.  Lastly, we show how the
boundedness of $t$ on the ``northern'' and ``southern'' boundaries
entails the boundedness of $t$ on all of ``northern'' and ``southern''
regions.  Thus, $t$ is globally bounded which gives

{\it Theorem 1.}
There exists an upper bound to the lengths of timelike curves in any
Tolman spacetime that possesses $S^3$ Cauchy surfaces and whose energy
density is positive.

This result is further strengthened by Theorem~5 where an explicit
expression for an upper bound is given in terms of initial data for
the spacetime on any spherically symmetric Cauchy surface.

The closed-universe recollapse conjecture has been investigated for
the Tolman spacetimes (through a variety of methods) and either
claimed to be true by argument \cite{ZG,wbB} or false by
counterexample \cite{HL}.  However, none of these analyses provide
either a true proof or counterexample of the closed-universe
recollapse conjecture for the Tolman spacetimes (in any reasonable
form).

Zel'dovich and Grishchuck showed that naive expectations regarding the
unbounded expansion of a closed Tolman spacetime that is ``locally
open'' are incorrect.  However, their work does not show that
there is indeed an upper bound to the lengths of timelike curves in
such a spacetime.  While their argument does show that the crossing of
flow lines will develop somewhere in the spacetime, it does not
necessarily mean that such an occurrence will be visible to all
observers.  Their work thus leaves open the possibility that some
observer may forever be unaffected by the crossing of flow lines.

Bonnor showed that his ``physically acceptable cosmological models''
must be everywhere ``elliptic'' and therefore must recollapse.  The
problem here is that the requirement for a Tolman spacetime to be
``physically acceptable'' is far too restrictive in that spacetimes
where the dust lines eventually intersect are ruled out.  Such
spacetimes are physically acceptable as long as the regions where such
crossing occur are excluded while keeping the spacetime globally
hyperbolic.  Even accepting the result, this still does not give a
finite upper bound on the lengths of causal curves as the time to
recollapse as one nears the ``poles'' may become infinite.

Finally, the counterexample to the closed-universe recollapse
conjecture offered by Hellaby and Lake is flawed in that it contains a
surface layer of matter that has associated with it a negative
pressure.  (This was noted by Bonnor \cite{wbB}.)\ \ In fact, a simple
example of the problem inherent with their spacetime is the following.
Consider the static spherically symmetric spacetime with $S^3$ Cauchy
surfaces with metric given by $g_{ab} = -(dt)_a(dt)_b + (dx)_a(dx)_b +
r^2 \Omega_{ab}$ where $r(t,x) = 1-|1-x|$, $\Omega_{ab}$ is the
standard unit-metric on the 2-sphere, and $0 \le x \le 2$, $-\infty
<t< \infty$.  (The ``poles'' are at $x=0,2$.)\ \  This spacetime is flat
everywhere except on the timelike surface $x=1$ where there is a
distributional stress-energy that possesses negative pressures.  So,
although this spacetime possesses $S^3$ Cauchy surfaces and has
infinite length timelike curves, since the energy conditions are not
met, it is not a counterexample to the closed-universe recollapse
conjecture as formulated above.

In Sec.~\ref{sec:basics} the basics of the spherically symmetric
spacetimes are presented.  The Einstein equations are given in a form
amenable to analysis, and the fields $r$ and $m$ are introduced and
their basic properties established.  Throughout this section the
analysis is quite general in that it is independent of the Tolman
matter assumption.  In Sec.~\ref{sec:tolman} the results obtained in
Sec.~\ref{sec:basics} are applied to the Tolman spacetimes and
Theorem~1 is proven.  Lastly, in Sec.~\ref{sec:dis} a few final
remarks are made regarding possible extensions of this work.

The conventions used herein are those of Ref.~\cite{rmW}.  In
particular, our metrics are such that timelike vectors have negative
norm and the Riemann and Ricci tensors are defined by $2
\nabla_{[a}\nabla_{b]} \omega_c = R_{abc}{}^d \omega_d$ and $R_{ab} =
R_{amb}{}^m$ respectively.  All metrics are taken to be $C^2$.  Our
units are such that $G = c = 1$.

\section{Basics} \label{sec:basics}

In this section, the Einstein equations for the spherically symmetric
spacetimes are presented and are used to prove

{\it Theorem 2.}
Fix any spherically symmetric spacetime $(M,g_{ab})$ that possesses $S^3$
Cauchy surfaces and that satisfies the non-negative-pressures and
dominant-energy conditions.  Then $m \ge 0$ and $r \le
\max_\Sigma(2m)$ where $\Sigma$ is any spherically symmetric Cauchy
surface for this spacetime.

In other words, the spheres of symmetry can't become arbitrarily large
and their associated ``quasi-local mass'' is always non-negative.

\subsection{Field Equations}

Recall that a spacetime $(M,g_{ab})$ is said to be {\it spherically
symmetric} if it admits the group $G \approx SO(3)$ of isometries,
acting effectively on $M$, each of whose orbits is either a two-sphere
or a point \cite{HE}.  The value of the non-negative scalar field $r$
at each $p \in M$ is defined so that $4\pi r^2$ is the area associated
with the orbit of $p$.  So, in particular, $r(p)$ is zero if the orbit
is a point, while $r(p)$ is positive if the orbit is a two-sphere.

For a spherically symmetric spacetime $(M,g_{ab})$ with $S^3$ Cauchy
surfaces, the set of points $p$ for which $r(p)=0$ (points whose
orbits are themselves) consists of two disconnected components which
we label $\gamma_n$ and $\gamma_s$.  It follows from the spherical
symmetry that these two sets must in fact be timelike geodesics.  So,
physically these curves are the world lines of the two privileged
observers for whom the universe actually appears spherically
symmetric.

As in the $S^1 \times S^2$ case, we can construct a two-dimensional
spacetime $(B,h_{ab})$ by setting $B=M/G$ (i.e.\ $B$ is the set of orbits)
and $h^{ab} = (\pi^* g)^{ab}$ where $\pi$ is the natural projection
map from $M$ to $M/G$.  However, unlike the $S^1 \times S^2$ case,
$(B,h_{ab})$ is a spacetime with boundary consisting of the two geodesics
$\pi(\gamma_n)$ and $\pi(\gamma_s)$.  (In other words, the boundary is
timelike with zero extrinsic curvature.  Further, it follows from the
fact that $(M,g_{ab})$ possesses $S^3$ Cauchy surfaces that $B \approx R
\times [0,1]$.)\ \  Because of this timelike boundary, the spacetime
$(B,h_{ab})$ is not globally hyperbolic in the traditional sense,
which makes working with the two-dimensional spacetime $(B,h_{ab})$
somewhat awkward.  It is for this reason that the basic results
concerning the spherically symmetric spacetimes shall be stated and
proved using the four-dimensional spacetime $(M,g_{ab})$.

Denote the projection operator onto the tangent space of each sphere
of symmetry by $q^a{}_b$.  (So, e.g., if $x^a$ is a vector tangent to
a surface of symmetry then $q^a{}_b x^b = x^a$, while if $x^a$ is
perpendicular to such a surface $q^a{}_b x^b = 0$.)\ \ From $q^a{}_b$ we
construct the projection operator onto the surfaces perpendicular to
the spheres of symmetry: $h^a{}_b = \delta^a{}_b - q^a{}_b$.  Thus,
using $g^{ab}$ and $g_{ab}$ to raise and lower the indices of these
tensor fields, we have the following decomposition of the metric
(where $r>0$)
\begin{equation}
g_{ab} = h_{ab} + q_{ab}.
\end{equation}

We further decompose $g_{ab}$ by decomposing $q_{ab}$.  Denote the
Killing vector fields associated with the action of $G$ on $M$ by
$\xi^a_\alpha$ where lower-case Greek indices are Lie-algebra indices.
With these we define the tensor field
\begin{equation} \label{defOmega}
\Omega^{ab} = \xi^a_\alpha \xi^b_\beta k^{\alpha\beta},
\end{equation}
where $k^{\alpha\beta}$ is the inverse of $k_{\alpha\beta} = -
\frac{1}{2} c^\mu{}_{\alpha\nu}c^\nu{}_{\beta\mu}$ (which is one-half
of the Killing-Cartan metric) and $c^{\gamma}{}_{\alpha\beta}$ are the
structure constants for the Lie algebra associated with $G$ (so that,
in particular, we have $[\xi_\alpha,\xi_\beta]^a =
c^{\gamma}{}_{\alpha\beta} \xi^a_\gamma$). This tensor has the
following properties: (1) It is tangent to the spheres of symmetry;
(2) It is spherically symmetric, ${\cal L}_{\xi_\alpha}
\Omega^{ab} = 0$; (3) If $\zeta^a$ is any spherically symmetric vector
field, then ${\cal L}_\zeta \Omega^{ab} = 0$; (4) On the spheres of
symmetry, $\Omega^{ab}$ is positive definite; (5) The area of any
sphere of symmetry computed using $\Omega^{ab}$ is $4\pi$; (6) On
$\gamma_n$ and $\gamma_s$, $\Omega^{ab} =0$.  These properties allow one
to think of $\Omega^{ab}$ as the preferred unit-metric on each sphere
of symmetry.

Define $\Omega_{ab}$ to be the inverse of $\Omega^{ab}$ so that
$\Omega^{am}\Omega_{mb} = q^a{}_b$ and $\Omega_{ab} = q^m{}_a q^n{}_b
\Omega_{mn}.$ We then have $q_{ab} = r^2 \Omega_{ab}$.  To show this,
consider the linear map $\Omega^{am}q_{mb}$.  By spherical symmetry
this map must be proportional to $q^a{}_b$ for otherwise its preferred
eigenvectors would violate the rotational symmetry about each point.
Thus, $\Omega^{ab}$ and $q^{ab}$ must be proportional.  Using the fact
that the area computed using $\Omega^{ab}$ is $4\pi$ and that the area
computed using $q^{ab}$ is $4 \pi r^2$ we find that $q^{ab} =
\frac{1}{r^2} \Omega^{ab}$ or more simply $q_{ab} = r^2 \Omega_{ab}$.

We thus arrive at the following decomposition of the metric $g_{ab}$
(where $r>0$)

\begin{equation}
g_{ab} = h_{ab} + r^2 \Omega_{ab}.
\end{equation}

Define $D_a$ to be that (torsion-zero) derivative operator associated
with the (unphysical) metric $h_{ab} + \Omega_{ab}$.  It then follows
(from arguments involving spherical symmetry) that $D_a h_{bc} = D_a
\Omega_{bc} = 0$.  Further, if $\omega_a$ is spherically symmetric
then $D_a \omega_b = h^m{}_a h^n{}_b \nabla_m \omega_n$, and if ${\cal
L}_\zeta \omega_a = 0$ for all spherically symmetric vector fields
$\zeta^a$ then $D_a \omega_b = q^m{}_a q^n{}_b \nabla_m \omega_n$.
These properties allow one to think of $D_a$ as the derivative
operator associated with $h_{ab}$ on the surfaces perpendicular to the
spheres of symmetry (or on $B$ if desired) and as the derivative
operator associated with $\Omega_{ab}$ on the spheres of symmetry.

We define the ``quasi-local mass'' $m$ by \cite{MS}
\begin{equation} \label{defm}
2m = r(1-D_m r D^m r).
\end{equation}
That this quantity deserves such a title is born out by its direct
relation to the stress-energy content of the spacetime (given by
Eq.~(\ref{D2m})), and in the vacuum case its being the mass of that
extended Schwarzschild spacetime to which this spacetime is locally
isometric.  Furthermore, $m$ is non-negative (proven in
Sec.~\ref{sec:m}) and is of great utility in all that follows.

Denote the scalar curvature associated with the surfaces perpendicular
to the spheres of symmetry by $R[h]$, and define $\epsilon^{ab}$ to be
either of the two antisymmetric tensor fields such that
$\epsilon^{ab}\epsilon^{cd} = -2 h^{a[c}h^{d]b}$.  With these
definitions, the Riemann tensor for the four-dimensional spacetime
$(M,g_{ab})$ is found to be
\begin{equation} \label{Riemann}
R_{abcd} = R[h] h_{c[a}h_{b]d} - {4 \over r} q_{[c|[a}D_{b]|}D_{d]}r +
{ 4m \over r^3} q_{c[a} q_{b]d}
\end{equation}
{}from which the Einstein tensor is computed to be
\begin{eqnarray} \label{Einstein}
G_{ab} = && {2 \over r} \left [ D_mD^m r - {m \over r^2} \right ]h_{ab} -
{2 \over r} D_a D_b r \nonumber \\
&& + \left [ {1 \over r} D_mD^m r - \case{1}{2} R[h] \right ] q_{ab}.
\end{eqnarray}

{}From Eq.~(\ref{Riemann}) we see that for the curvature to be finite
on $\gamma_n$ and $\gamma_s$ then $m/r^3$ must be finite on each of
these two curves.  Thus, both $m/r^2$ and $m/r$ must be zero on
$\gamma_n$ and $\gamma_s$ from which it follows that $D^a r$ is
unit-spacelike on $\gamma_n$ and $\gamma_s$.

Using the above decomposition of the metric $g_{ab}$, we decompose the
Einstein equation as follows.  Decompose the Einstein tensor $G_{ab}$
(which through Einstein's equation is proportional to the
stress-energy tensor of the matter) into the pair of spherically
symmetric fields $(\tau^{ab},P)$ where $\tau^{ab}$ is the purely
``radial'' part of the Einstein tensor ($\tau^{ab}=h^a{}_m h^b{}_n
G^{mn}$) and $P$ is the pressure associated with the spheres of
symmetry ($P = {1 \over 2} G^{mn} q_{mn}$).  Then,
\begin{equation}
G_{ab} = \tau_{ab} + P q_{ab}.
\end{equation}
The twice-contracted Bianchi identity ($\nabla_b G^{ab} = 0$) requires
that the pair $(\tau^{ab},P)$ satisfy
\begin{equation} \label{Bianchi}
D_b(r^2 \tau^{ab}) = P D^a r^2.
\end{equation}

With this decomposition of $G_{ab}$, Eq.~(\ref{Einstein}) becomes the
pair of equations
\begin{eqnarray}
D_a D_b r  = && {m \over r^2} h_{ab} -
{r \over 2} \tau^{mn}\epsilon_{ma}\epsilon_{nb}, \label{DDr} \\
R[h] = && {4m \over r^3} + (\tau_m{}^m - 2P). \label{Rh}
\end{eqnarray}
{}From Eq.~(\ref{DDr}) we arrive at the following simple and very
useful equation relating the gradient of $m$ algebraically to the
stress-energy tensor

\begin{equation} \label{D2m}
D_a(2m) = r^2 \tau^{mn} \epsilon_{ma}\epsilon_{nb} D^br.
\end{equation}
If desired, one could work with Eqs.~(\ref{Bianchi}--\ref{D2m}) as
equations on the two-dimensional spacetime $(B,h_{ab})$ as was done in
the $S^1 \times S^2$ case.

\subsection{Non-negativity of $m$} \label{sec:m}

In the study of the spherically symmetric spacetimes with $S^1 \times
S^2$ Cauchy surfaces, it was shown that (under certain energy
conditions) the ``quasi-local mass'' $m$ is globally bounded from
below by a positive constant.  That is, not only is $m$ non-negative
in that case, it can neither become zero nor get arbitrarily close to
zero.  In the $S^3$ case, however, $m$ is zero on the curves
$\gamma_n$ and $\gamma_s$.  So, the best global bound one can hope for
in this case is for $m$ to be non-negative.  In fact, as we show
below, not only is $m$ non-negative, but in fact $m$ can be zero only
in flat (hence vacuum) regions about $\gamma_n$ or $\gamma_s$.  It may
be noted that the argument used here to establish the non-negativity
of $m$ is an ``initial value'' type of argument in that we show that
$m$ is non-negative on any Cauchy surface from which it follows that
$m$ is non-negative everywhere.

We begin with

{\it Lemma 1.}
Fix any spherically symmetric globally hyperbolic spacetime $(M,g_{ab})$
that satisfies the dominant-energy condition.  Fix a spherically
symmetric Cauchy surface $\Sigma$ therein and let $C$ be any
spherically symmetric compact subset of $\Sigma$. Then
\begin{equation} \label{lowbound}
\min_C (2m) \ge \min(\min_{\partial C} (2m),\min_C (r)).
\end{equation}

{\it Proof.}
Denote the right-hand side of Eq.~(\ref{lowbound}) by $2\mu$ and
consider the open proper subset $U$ of $C$ defined by $U=\{ s \in
C|2m(s) < 2\mu\}$.  On $U$, $2m<r$ so $D^a r$ is necessarily
spacelike.  Denote, by $s^a$, the unit spherically symmetric vector
field on each connected component of $U$, tangent to the surface
$\Sigma$, such that $s^aD_a r >0$.  Then, by Eq.~(\ref{D2m}) and the
fact that $\tau^{ab}$ satisfies the dominant-energy condition, we have
$s^aD_a(2m) \ge 0$ on $U$.  Using this fact and noting that $2m=2\mu$
on the boundary of $U$, we conclude that $2m=2\mu$ on $U$.  This
contradicts the definition of $U$, so $U$ must be empty.  This
establishes the above lower bound for $m$ on $C$.$\Box$

The non-negativity of $m$ in the $S^3$ case follows as a simple
consequence of this result.  For any point $p \in M$, let $\Sigma$ be
any spherically symmetric Cauchy surface for $(M,g_{ab})$ with $p \in
\Sigma$.  Take $C=\Sigma$.  Since $\partial \Sigma$ is empty and
$\min_\Sigma (r) = 0$, by Lemma~1, $m \ge 0$ on $\Sigma$.  So, $m$ is
non-negative at $p$.

In fact, we can strengthen this last result in that if $S$ is a sphere
of symmetry for which $m(S)=0$, then $m=0$ for all points ``inside''
of $S$.  For a sphere of symmetry $S$ with $D^a r$ spacelike (i.e.\
the sphere is neither future/past trapped nor marginally trapped), we
shall say that a point $p\in M$ is {\it inside} of $S$ if $p$ lies in
the component of $(J^+(S)\cup J^-(S))^c$ for which $D^a r$ is outward
pointing.  (For $(M,g_{ab})$ globally hyperbolic and connected, the
set $(J^+(S)\cup J^-(S))^c$ can have at most two components and may
have only one as in the $S^1 \times S^2$ case.)\ \ With this
definition, we have

{\it Theorem 3.}
Fix any spherically symmetric globally hyperbolic spacetime $(M,g_{ab})$
that satisfies the dominant-energy condition.  If $m(S)=0$ for some
sphere of symmetry $S$, then $m=0$ for all points inside of $S$.

{\it Proof.}
Since $m(S)=0$, by Eq.~(\ref{defm}) we have $D_ar D^a r = 1$ on $S$
showing that $D^a r$ is spacelike on $S$.  For any point $p$ inside of
$S$, let $\sigma$ be a radial spacelike curve from $S$ to $p$ with
tangent vector $s^a$.  (Such a curve always exists.)\ \ Let $\sigma'$
be the maximal subset of $\sigma$ connected to $S$ on which $D^a r$ is
spacelike.  Using the fact that $s^a D_a r < 0$ on $\sigma'$, the fact
that $\tau^{ab}$ satisfies the dominant energy condition, and
Eq.~(\ref{D2m}), we find $s^aD_a(2m) \le 0$ on $\sigma'$.  Thus, for
all points $q \in \sigma'$ we have the inequalities $0 \le m(q) \le
m(S) = 0$ showing that $m=0$ on $\sigma'$.  However, this shows that
$D^a r D_a r = 1$ on $\sigma'$ so in fact $\sigma'=\sigma$.  Thus,
$m(p)=0$.$\Box$

It is in this sense (a sense as strong as one could hope for) that $m$
can vanish only ``about'' the poles.  Further, such regions must be
flat.

{\it Theorem 4.}
Fix any spherically symmetric globally hyperbolic spacetime $(M,g_{ab})$
that satisfies the dominant-energy condition.  If $m(S)=0$ for some
sphere of symmetry $S$ in $M$, then the spacetime inside of $S$ is flat
($R_{abcd}=0$.)

{\it Proof.}
Denote the spacetime inside of $S$ by ${\cal F}$.  By
Eq.~(\ref{Riemann}), a spherically symmetric spacetime is flat on an
open set ${\cal F}$ iff $m=0$, $D_aD_b r = 0$, and $R[h]=0$ on ${\cal
F}$.  Since $m(S)=0$, by Theorem~3, $m=0$ on ${\cal F}$.  So, by
Eq.~(\ref{D2m}), $\tau^{mn}(\epsilon_{ma}D^a r)(\epsilon_{nb}D^b r)=0$
on ${\cal F}$.  But, since $\epsilon_{am}D^m r$ is timelike, by the
dominant-energy condition, it must be the case that $\tau^{ab}=0$ and
consequently $P=0$ on ${\cal F}$. So, by Eq.~(\ref{DDr}), we have
$D_aD_b r = 0$ and by Eq.~(\ref{Rh}), $R[h]=0$.$\Box$

Likewise, a region that is vacuum about a pole must be flat since, by
Eq.~(\ref{D2m}), in that region $m$ is zero and thus must be flat, by
Theorem~4.  However, it is interesting to note that a spherically
symmetric spacetime with $S^3$ Cauchy surfaces cannot be completely
vacuum since, by Eq.~(\ref{defm}), $m$ is manifestly positive where
$r$ reaches its maximum value on any Cauchy surface which, by
Eq.~(\ref{D2m}), demands that $G^{ab}$ be non-zero somewhere on that
surface.

\subsection{Upper bound for $r$}

We now establish the upper bound for $r$ given in Theorem~2.  First,
we have the following upper bound for $r$ on a Cauchy surface.

{\it Lemma 2.}
Fix any spherically symmetric globally hyperbolic
spacetime $(M,g_{ab})$ that possesses compact Cauchy surfaces and that
satisfies the dominant-energy condition.  For any spherically
symmetric Cauchy surface $\Sigma$ of
$(M,g_{ab})$ we have
\begin{equation}
\max_\Sigma (r) \le \max_\Sigma(2m).
\end{equation}

{\it Proof.}
Consider a point $p$ where $r$ reaches its maximum value on $\Sigma$.
At such a point $D^a r$ is necessarily timelike or zero.  Hence
\begin{equation}
\max_\Sigma (r) = r(p) \le 2m(p) \le \max_\Sigma (2m),
\end{equation}
where the first inequality is by Eq.~(\ref{defm}).$\Box$

Next, we have the following global upper bound for $r$.

{\it Lemma 3.}
Fix any spherically symmetric globally hyperbolic spacetime $(M,g_{ab})$
that satisfies the non-negative-pressures and dominant-energy
conditions.  For any spherically symmetric Cauchy surface $\Sigma$ of
$(M,g_{ab})$ we have
\begin{equation}
r \le \max(\sup_\Sigma (r), \sup_\Sigma(2m)).
\end{equation}

{\it Proof.}
It suffices, since we can always reverse the roles of past and future,
to establish this bound for any $p\in D^+(\Sigma)$.

Consider any point $q$, where $r$ reaches its maximum value on the
compact set $C=J^-(p) \cap D^+(\Sigma)$.  If $q\in C\cap\Sigma$, then
$r(p) \le r(q) \le \sup_\Sigma (r)$.  If $q \notin C \cap \Sigma$,
then $D^a r$ must be either past-directed timelike, past-directed
null, or zero, at $q$, for otherwise there would exist a past-directed
timelike direction along which $r$ would increase.  We now show, in
all three cases, that $r(q) \le \sup_\Sigma (2m)$.

If $D^a r$ is past-directed timelike or past-directed null at $q$,
then, by Eq.~(\ref{defm}), $r(q) \le 2m(q)$.  Consider the maximal
integral curve $\sigma$ of $D^a r$ with future end point $q$ on which
$D^a r$ is past-directed timelike or past-directed null.  The curve
$\sigma$ does not have a past endpoint.  [Proof: Using the
non-negative-pressures condition and Eq.~(\ref{DDr}) we have $(D^a
r)D_a(D_m r D^m r) \le 0$, on $\sigma$, showing that for $\sigma$ to
have a past endpoint $q'$, then $D^a r$ must be past-directed null all
along $\sigma$ and zero at $q'$.  In this case, since $\sigma$ is
radial and null, it must be a geodesic curve.  Affinely parameterize
$\sigma$ by $\lambda$ and denote its associated tangent vector by
$k^a$.  Then, using the null-convergence condition ($G_{ab}k^ak^b \ge
0$ for all null $k^a$, which follows from such energy conditions as
the dominant-energy condition) and Eq.~(\ref{DDr}), we find that $d^2
r/d \lambda^2 \le 0$.  But, since $dr/d\lambda < 0$ at $q$, it is
impossible for $dr/d\lambda$ and hence $D^a r$ to be zero at $q'$.
Thus, a past endpoint $q'$ cannot exist.]\ \ So, since $\sigma$ is a
past-directed causal curve with future endpoint $q \in D^+(\Sigma)$
and without past endpoint, it must, by global hyperbolicity, intersect
$\Sigma$.  Again using the non-negative-pressures condition and
Eq.~(\ref{defm}), we find that $(D^a r)D_a(2m) \ge 0$, on $\sigma$, so
that $r(q) \le 2m(q)
\le 2m(\sigma \cap \Sigma) \le \sup_\Sigma(2m)$.

If $D^a r$ vanishes at $q$, then so does $D_a(-D_m r D^m r)$.  Using
Eq.~(\ref{DDr}) we find that, at $q$, for radial unit past-directed
timelike $t^a$,
\begin{eqnarray} \label{DDnorm}
t^at^bD_aD_b(-D_m r D^m r) = && {1 \over 2r^2} +
\tau^{mn}\epsilon_{ma}\epsilon_{nb}t^at^b \nonumber \\
&& + {r^2 \over 2} (\tau^{mn} \epsilon_{ma}t^a)(\tau^{pq}
\epsilon_{pb}t^b)h_{nq}.\nonumber \\
\end{eqnarray}
The first term is manifestly positive; the second term is non-negative
by the non-negative-pressures condition; and by the dominant-energy
condition, there exist $t^a$ for which the last term is non-negative.
[Sketch of proof: Use the fact that $h^{ab}=-2k^{(a}l^{b)}$ where
$k^a$ and $l^b$ are two linearly-independent radial past-directed null
vectors.  Set $u^a = - \tau^a{}_b k^b$ and $v^a = - \tau^a{}_b l^b$.
If either $u^a$ or $v^a$ is timelike (necessarily being past-directed)
then taking $t^a$ to be colinear with either (timelike) vector
guarantees that the last term is zero.  Otherwise, if $u^a$ and $v^a$
are null or zero, then any $t^a$ will do.]\ \ Consider the
past-directed timelike geodesic starting at $q$, with initial tangent
vector $t^a$ such that the last term, in Eq.~(\ref{DDnorm}), is
non-negative. Then, at $q$, $t^aD_a(t^bD_b(-D_mrD^mr)) > 0$, and, by
the non-negative-pressures condition, $t^aD_a(t^bD_b r) < 0$.  Hence,
$D^a r$ immediately becomes past-directed timelike along the curve.
{}From our analysis of such points, we conclude that $r(q) = 2m(q) \le
\sup_\Sigma(2m)$.$\Box$

Combining Lemmas~2 and~3, the global upper bound $r \le \max_\Sigma
(2m)$ is established.

\section{Tolman Result} \label{sec:tolman}

The Tolman spacetimes are the spherically symmetric spacetimes whose
matter content is dust (a perfect fluid with zero pressure) \cite{LL}.
The Einstein equation is then
\begin{equation}\label{stressenergy}
G_{ab} = 8 \pi \rho u_a u_b
\end{equation}
where $\rho$ is the energy-density of the dust and $u^a$ is a
(spherically symmetric) unit future-directed timelike vector field.
Decomposing $G_{ab}$, as described in Sec.~II, we have $\tau_{ab}=
8\pi \rho u_a u_b$ and $P=0$.  Thus, by Eq.~(\ref{Bianchi}), we find
that $u^a$ is geodetic and that $D_a(\rho r^2 u^a) = 0$.  For a Tolman
spacetime to satisfy the dominant-energy and non-negative-pressures
conditions it is necessary and sufficient that $\rho$ be non-negative.

Strictly speaking, $u^a$ need only be defined where $\rho$ is
positive.  It is for this reason that we haved restricted ourselves to
the nowhere empty Tolman spacetimes.  However, the more general case
where $\rho$ can vanish in a region (and so $u^a$ is undefined there)
can be analyzed if $u^a$ can be extended to these regions in such a
way that it is spherically symmetric and geodetic.  However, whether
this extension can be accomplished in general is unclear, though it
seems likely.  If it can be accomplished, then all the results that
follow also apply for these spacetimes.

One problem with dust as a matter source is that dust lines may cross
thus destroying the dust assumption embodied in
Eq.~(\ref{stressenergy}).  In this work we shall adopt a strict
definition of a Tolman spacetime by demanding that there are no such
crossings.  So, for example, if one where to evolve one of these
spacetimes from a given initial data surface and such a crossing were
to develop, then the evolution is to be stopped where the crossing
occurs and continued elsewhere only to the extent that the constructed
spacetime is globally hyperbolic.

One feature of the Tolman spacetimes that makes their study tractable
is that $m$ is constant along the integral curves of $u^a$.  For
proof, using Eq.~(\ref{D2m}) we have $u^aD_a(2m) = (8\pi\rho u^m
u^n)(\epsilon_{ma}u^a)(\epsilon_{nb}D^br) = 0$.  Further, by
Theorem~2, $r$ is bounded from above by
\begin{equation} \label{defrM}
r_M = \max_\Sigma(2m),
\end{equation}
where $\Sigma$ is any spherically symmetric Cauchy surface for this
spacetime.  Using these facts we have

{\it Lemma 4.}
Any integral curve $\gamma$ of $u^a$ for which $m(\gamma)>0$ is
bounded in length by
\begin{equation}\label{bound}
T[\gamma] = \pi \sqrt{r_M^3 \over 2m(\gamma)}.
\end{equation}

{\it Proof.}
Since $m(\gamma)>0$, $r$ must be strictly positive on
$\gamma$ and so $0 < r \le r_M$ on $\gamma$.  Further, since $\gamma$
is geodetic it follows that
\begin{equation}\label{d2rdt2}
{d^2r \over dt^2} = u^au^bD_aD_b r = - {m(\gamma) \over r^2},
\end{equation}
where the second equality follows from Eq.~(\ref{DDr}) and the
vanishing of the pressure
$\tau^{mn}(\epsilon_{ma}u^a)(\epsilon_{nb}u^b)$.  However, it is a
straightforward exercise to show that it is impossible to satisfy
Eq.~(\ref{d2rdt2}) and the inequalities $0 < r \le r_M$ for a time
$T[\gamma]$ or greater.$\Box$

Thus, for the Tolman spacetimes, we quite easily arrive at the result
that all of integral curves of the fluid flow (with $m>0$) must be
incomplete.  However, this is quite far from showing that there is an
upper bound to the lengths of all the timelike curves in such a
spacetime.  For, consider the upper bound given by Eq.~(\ref{bound}).
As we approach either $\gamma_n$ or $\gamma_s$, $m(\gamma)$ approaches
zero so that, at this stage in the argument, it is still conceivable
that while curves ``between the poles'' will be finite in length,
$\gamma_n$, $\gamma_s$, and ``nearby'' curves could be infinite in
length!  This is by no means a minor technicality.  For instance, it
is conceivable that integral curves of $u^a$ initially near $\gamma_n$
``peel away'' in such a way as to hide their demise from $\gamma_n$.

While this ``peeling'' of neighboring curves from $\gamma_n$ ($\gamma_s$)
may occur, they can't do it in such a way that their demise can't be
seen by $\gamma_n$ ($\gamma_s$).  Or, said better, we will see that an
integral curve of $u^a$ sufficiently near $\gamma_n$ ($\gamma_s$) having
a finite length requires that $\gamma_n$ ($\gamma_s$) have a finite
length (Lemma~7 (8)).

\subsection{The ``time function'' $t$}

It follows directly from the facts that the vector field $u^a$ is
spherically symmetric, unit, and geodetic that $D_a u_b =
(D_mu^m)(h_{ab}+u_au_b)$ from which it is apparent that $u_a$ is
closed: $(du)_{ab}=0$.  Thus, since $M$ is simply connected, there
exists a scalar field $t$ (unique up to the addition of a constant)
such that
\begin{equation} \label{dt}
u_a = - (dt)_a.
\end{equation}
The ``time function'' $t$ is a great aid in establishing an upper
bound to the lengths of timelike curves in the Tolman spacetimes.
Noting that $u^a(dt)_a = 1$ we see, by Lemma~4, that $t$ is bounded
{}from above and below along any integral curve of $u^a$ for which $m$
is positive.

Those familiar with the Tolman spacetimes may recognize $t$ as being
one of the coordinates in which the Tolman spacetimes are usually
presented.  In fact, if a hypersurface everywhere orthogonal to $u_a$
exists, then $t$ is one of the synchronous (Gaussian normal)
coordinates associated with this hypersurface and the geodetic vector
field $u^a$ \cite{MTW}.  However, since there is no guarantee that
such a surface will exist, the above construction of $t$ is preferred.
(We further caution the reader that some or all surfaces of constant
$t$, though spacelike, may not be Cauchy surfaces.)

Recall that the distance function $d(p^-,p^+)$ is defined for $p^+ \in
I^+(p^-)$ to be the least upper bound of the lengths of timelike
curves from $p^-$ to $p^+$ (and to is defined to be zero otherwise)
\cite{HE}.  Thus, the least upper bound to the lengths of timelike
curves in a spacetime $(M,g_{ab})$ will be finite iff $d(p^-,p^+)$ is
bounded for all $p^\pm \in M$.

For any two points $p^\pm \in M$ with $p^+ \in I^+(p^-)$, let $\mu$ be
a future directed timelike curve connecting $p^-$ to $p^+$ and having
length $d(p^-,p^+)$.  (In other words, $\mu$ is a maximal geodesic
connecting the two points.)\ \ Parameterize $\mu$ by $\tau$ so that
its tangent vector $v^a$ has unit norm.  We then have
\begin{eqnarray} \label{dtt}
d(p^-,p^+) = && \int_\mu d\tau =
\int_\mu {dt \over (v^a(dt)_a)} \nonumber \\
\le && \int_\mu dt = t(p^+)- t(p^-).
\end{eqnarray}
The second equality follows from the fact that $dt/d\tau = v^a(dt)_a$.
The inequality follows from the fact that $v^a(dt)_a = -v^au_a$ is no
less than unity as both $v^a$ and $u^a$ are unit future-directed
timelike vectors.  So, to bound the length of a timelike curves, we
need only find a global bound on the difference $t(p^+)-t(p^-)$.

\subsection{Bounding $t$}

Denote the one-parameter local pseudo group of diffeomorphisms
associated with the vector field $u^a$ by $\exp_t \colon M \rightarrow
M$ \cite{CDD}.  For any set $C$ define $\exp_\pm(C)$ be the set of
those $p \in M$ such that $p = \exp_s(c)$ for some $c \in C$ and some
$\pm s \ge 0$.  Set $\exp(C)$ to be the union of these two sets.  So,
in particular for a point $p\in M$, $\exp(p)$ is the integral curve of
$u^a$ that passes through the point $p$ while $\exp(p)\cap\Sigma$ is
the point where this integral curve intersects $\Sigma$.

To establish bounds on $t$, we divide the spacetime $(M,g_{ab})$ into
three regions.  We do this by first dividing up the Cauchy surface
$\Sigma$.  Let $C_n$ and $C_s$ be any two spherically symmetric
compact connected subsets of $\Sigma$ such that
\begin{enumerate}
\item[(i)] $C_n$, $C_s$ intersects $\gamma_n$, $\gamma_s$ respectively,
\item[(ii)] $D^a r$ is spacelike on $C_n$ and $C_s$,
\item[(iii)] $m[\partial C_n], m[\partial C_s] > 0$.
\end{enumerate}
Note that such sets always exist and that they are always disjoint.
Define $C_m$ to be the closure of $\Sigma - (C_n \cup C_s)$ in
$\Sigma$.  (Note that $\partial C_m = \partial C_n \cup \partial
C_s$.)\ \ Using the three sets $C_m$, $C_n$, and $C_s$, we divide the
full spacetime $M$ into three regions: a ``middle'' region,
$\exp(C_m)$; a ``northern'' region, $\exp(C_n)$; and a ``southern''
region, $\exp(C_s)$.  Denote the boundaries of these regions (being
timelike three-surfaces with $u^a$ tangent thereto) by ${\cal T}_n =
\partial \exp(C_n) = \exp(\partial C_n)$ and ${\cal T}_s = \partial
\exp(C_s) = \exp(\partial C_s)$.

\subsubsection{Bounds on the middle region} \label{sec:boundM}

Setting
\begin{equation} \label{TM}
T_m = \pi \sqrt{r_M^3 \over \min_{C_m}(2m)},
\end{equation}
we have

{\it Lemma 5.}
For any $p^\pm \in \exp_\pm(C_m)$ there are $a^\pm \in C_m$ such that

\begin{equation} \label{difftM}
\pm t(p^\pm) < \pm t(a^\pm)+ T_m.
\end{equation}

{\it Proof.} Setting $a^\pm = \exp(p^\pm) \cap \Sigma \in C_m$ this is
a simple application of Lemma~4 as $p^\pm$ and $a^\pm$ lie on the same
integral curve of $u^a$ with $2m \ge \min_{C_m}(2m)$.  (Lemma~1 gives
a positive lower bound for this last quantity.) $\Box$

\subsubsection{Bounds on the northern region} \label{ref:boundN}

To obtain bounds on $t$ in the region $\exp(C_n)$, we first define
${}^\pm k^a$ to be the spherically symmetric null vector fields such
that ${}^\pm k^a (dt)_a = \pm 1$ and ${}^\pm k^a D_a r = +1$ on
$\gamma_n$.  (Actually, ${}^\pm k^a$ are not well-defined on $\gamma_n$.
However, this is a mere nuisance and not a fundamental problem.)\ \
Set
\begin{equation}\label{kappaN}
\kappa_n^\pm = \min_{C_n}({}^\pm k^a D_a r) > 0.
\end{equation}

{\it Lemma 6.}
${}^\pm k^aD_a r \ge \kappa_n^\pm$ on $\exp_\mp(C_n)$.

{\it Proof.}
Consider
\begin{equation}\label{tDkDr}
u^aD_a ({}^\pm k^bD_b r) = u^a \; {}^\pm k^b D_aD_b r = \mp {m \over r^2},
\end{equation}
where the first equality follows from the fact that $u^aD_a {}^\pm k^b
= 0$ and the second from the fact that ${}^\pm k^au_a = \mp 1$ and the
vanishing of the pressure for these spacetimes
($\tau^{mn}(\epsilon_{ma}u^a)\epsilon_{nb} = 0$).  Set $Q^\pm = {}^\pm
k^aD_a r$.  For any point $p \in \exp_+(C_n)$, there is a unique
integral curve $\gamma$ of $u^a$ from $C_n$ to $p$.  But, by
Eq.~(\ref{tDkDr}), the quantity $Q^-$ is non-decreasing along
$\gamma$.  Thus, $Q^-(p) \ge Q^-(\gamma \cap C_n) \ge \kappa_n^-$.
Likewise, for any point $p \in \exp_-(C_n)$, there is a unique
integral curve $\gamma$ of $u^a$ from $p$ to $C_n$.  But, by
Eq.~(\ref{tDkDr}), the quantity $Q^+$ is non-increasing along
$\gamma$.  Thus, $Q^+(p)
\ge Q^+(\gamma \cap C_n) \ge \kappa_n^+$.$\Box$

With this technical lemma we have

{\it Lemma 7.}
For any $p^\pm \in \exp_\pm(C_n)$ there are $a^\pm \in C_n$ such that
\begin{equation}\label{difftN}
\pm t(p^\pm) < \pm t(a^\pm) + T_m + {r_M \over \kappa_n^\mp}.
\end{equation}

{\it Proof.}
Set $q^\pm = \lambda^\mp \cap \partial(\exp_\pm(C_n))$ where
$\lambda^\mp$ is the integral curve of ${}^\mp k^a$ starting from
$p^\pm$.  Consider
\begin{eqnarray}
r(q^\pm) - r(p^\pm) = &&
\int_{\lambda^\mp} dr =
\int_{\lambda^\mp} ({}^\mp k^a(dr)_a)(\mp dt) \nonumber \\
\ge && \kappa_n^\mp \int_{\lambda^\mp} (\mp dt) \nonumber \\
= && \mp \kappa_n^\mp (t(q^\pm) - t(p^\pm)).
\end{eqnarray}
The second equality follows from the facts that ${}^\mp k^a$ is
tangent to the curve $\lambda^\mp$ and that ${}^\mp k^a (dt)_a = \mp
1$.  This together with the inequality $r_M \ge r(q^\pm)-r(p^\pm)$
gives us
\begin{equation}
\pm t(p^\pm) \le \pm t(q^\pm) + {r_M \over \kappa_n^\mp}.
\end{equation}
Now, if $q^\pm \in C_n$ then taking $a^\pm = q^\pm$ Eq.~(\ref{difftN})
follows.  Otherwise, if $q^\pm \in {\cal T}_n$ then take $a^\pm =
\exp(q^\pm)\cap C_n$.  Then Eq.~(\ref{difftN}) follows from the fact
that
\begin{eqnarray}
\pm t(q^\pm) = &&  \pm t(a^\pm) \pm (t(q^\pm) - t(a^\pm)) \nonumber \\
< && \pm t(a^\pm) + T_m.
\end{eqnarray}
This last inequality follows from again applying Lemma~4 and the fact
that $2m({\cal T}_n) = 2m(\partial C_n) \ge \min_{C_m}(2m)$.$\Box$

\subsubsection{Bounds on the southern region} \label{sec:boundS}

Now define ${}^\pm k^a$ to be the null vector fields such that ${}^\pm
k^a (dt)_a = \pm 1$ and ${}^\pm k^a D_a r = +1$ on $\gamma_s$.  Set
\begin{equation}\label{kappaS}
\kappa_s^\pm = \min_{C_s}({}^\pm k^a D_a r) > 0.
\end{equation}
Then, by exactly the same methods as above we have

{\it Lemma 8.}
For any $p^\pm \in \exp_\pm(C_s)$ there are $a^\pm \in C_s$ such that
\begin{equation}\label{difftS}
\pm t(p^\pm) < \pm t(a^\pm) + T_m + {r_M \over \kappa_s^\mp}.
\end{equation}

\subsection{Bound on the Lengths of Timelike Curves}

With the bounds established in Secs.
\ref{sec:boundM}--\ref{sec:boundS}, we now prove Theorem~1 by
proving

{\it Theorem 5.}
In any Tolman spacetime that possesses $S^3$ Cauchy surfaces and whose
energy density is positive
\begin{equation}\label{grandbound}
d(p^-,p^+) < 2T_m + \tau^+ + \tau^- + \Delta t,
\end{equation}
where $T_m$ is given by Eq.~(\ref{TM}), $\tau^\pm$ and $\Delta t$
are given by
\begin{eqnarray}
\tau^\pm = && \max \left ( {r_M \over \kappa^\pm_n},{r_M \over
\kappa^\pm_s} \right ), \label{tau} \\
\Delta t = && \max_{a,b\in\Sigma}(t(b)-t(a)) \label{Delta t},
\end{eqnarray}
$\kappa^\pm_n$ and $\kappa^\pm_s$ are given by Eqs.~(\ref{kappaN}) and
(\ref{kappaS}) respectively, and $r_M$ is given by Eq.~(\ref{defrM}).

{\it Proof.}
Without loss in generality, taking $p^\pm \in D^\pm(\Sigma)$ it
follows directly from Eq.~(\ref{dtt}) and Lemmas~5, 7, and~8 that
\begin{eqnarray}
d(p^-,p^+) \le && t(p^+) - t(p^-) \nonumber \\
< && 2T_m + \tau^+ + \tau^- + (t(a^+)-t(a^-))
\end{eqnarray}
for some $a^\pm \in \Sigma$.  Eq.~(\ref{grandbound}) now follows from
the fact that $(t(a^+)-t(a^-)) \le \Delta t$.$\Box$

All the quantities that appear in the upper bound given by
Eq.~(\ref{grandbound}) are calculable from the initial data on an
initial data surface.  Once $C_n$ and $C_s$ are chosen, that $T_m$ and
$\tau^\pm$ can be calculated is clear.  Further, the quantity $\Delta
t$ can be calculated since $t(b)-t(a) = - \int_a^b u_m$ where the
integral is along any path from $a$ to $b$ in the Cauchy surface.
(Note that if $\Sigma$ is everywhere normal to $u_m$, then $\Delta
t=0$.  However, as mentioned earlier, such a surface may not exist!)

As a check on the bound given by Theorem~5, we use the $k=+1$
Friedmann-Robertson-Walker spacetimes with dust as a source.  For
these spacetimes it is a straightforward exercise to show that the
least upper bound to $d(p^-,p^+)$ is $\pi \max_\Sigma(2m)$ where
$\Sigma$ is any Cauchy surface.  Since $T_m > \pi \max_\Sigma(2m)$,
the inequality given by Eq.~(\ref{grandbound}) does hold for these
spacetimes.

Lastly, we note that the bound given by Theorem~5 may be a bit weaker
than can be argued.  As evidence for this, it is not too difficult to
show that for $p^\pm \in \exp(C_m)$ that $d(p^-,p^+) < T_m$.  (The
idea is to show that for such points, the geodesic $\mu$ attaining the
length $d(p^+,p^-)$ will remain in $\exp(C_m)$ and thus on $\mu$, we
have $2m \ge \min_{C_m}(2m)$.  Using these facts it follows by the
argument used in Ref.~\cite{gaB} that $d(p^-,p^+) < T_m$.)

\section{discussion} \label{sec:dis}

Now that we have a proof of the closed-universe recollapse conjecture
for the Tolman spacetimes, we ask whether its proof can be
generalized.  After all, such a result would be of far greater
interest than the more limited Tolman result.  Unfortunately, too many
properties of the Tolman spacetimes are used in the proof to make any
generalizations apparent.  First, how does one generalize the
``time-function'' $t$?  Even for a perfect fluid, where there is a
preferred vector field $u^a$, a similar construction of a
``time-function'' fails as $u_a$ fails to be closed.  One possible
generalization might involve the construction of a (locally defined)
geodetic vector field.  Then, as for Tolman, a (locally defined)
``time-function'' would exist.  Yet, how such a construction might
arise is not clear.  Second, even if such a geodetic congruence were
to be constructed, the proof of Lemma~6 would not go through.  With
the presence of a pressure, the right-hand side of Eq.~(\ref{tDkDr})
does not have any definite sign.  In light of these difficulties, for
the time being, the Tolman result presented here will remain the only
proof of the closed-universe recollapse conjecture for a class of
spacetimes that are spherically symmetric, have $S^3$ Cauchy surfaces,
and are spatially inhomogeneous.

Lastly, we offer the following piece of evidence, taking the form of a
gedanken experiment, which offers hope that the closed-universe
recollapse conjecture is true for the spherically symmetric spacetimes
with $S^3$ Cauchy surfaces.  Imagine two observers, one at $\gamma_n$,
the other at $\gamma_s$, each surrounded by a spherically symmetric
space station of mass $\mu$ designed to protect them from any outside
dangers--e.g.,\ infalling matter.  (The argument, as currently
formulated, requires both observers, but it seems plausible that one
such observer is sufficient for the argument to go through.)\ \ Can
such observers be protected forever, and therefor live forever, using
such an arrangement?  For simplicity, suppose we demand that their
protection is so good that the spacetime inside the space stations is
static.  In that case consider the longest timelike curve connecting
the outer surface of one of the stations to itself at two different
times.  Such a curve can be shown to lie completely in the region
between the two stations.  Further, it can be shown that in this
region $m$ is bounded from below by a positive constant (no greater
than $\mu$).  It follows from the same argument used to place an upper
bound on the lengths of timelike curves in the $S^1 \times S^2$
spherically symmetric spacetimes that there is an upper bound to the
length of such a curve connecting the outer surface of the station at
two different times.  Thus, the length of time such an arrangement can
be maintained (being less than this upper bound) is finite.
Unfortunately, how this scenario can be parlayed into a more general
theorem is not clear.

\end{document}